\documentclass[a4paper,11pt]{article}
\usepackage{pos}

\title{Finite-temperature critical point of heavy-quark QCD on large lattices}
\ShortTitle{Critical point of heavy-quark QCD}

\author*[a]{Kazuyuki Kanaya}
\author[b]{Ryo Ashikawa}
\author[c]{Shinji Ejiri}
\author[d,e]{Masakiyo Kitazawa}
\author[c]{Hiroto Sugawara}

\onbehalf{for the WHOT-QCD collaboration}

\affiliation[a]{Tomonaga Center for the History of the Universe, University of Tsukuba,\\
Tsukuba, Ibaraki 305-8571, Japan}

\affiliation[b]{Department of Physics, Osaka University,\\
Osaka, Osaka 560-0043, Japan}

\affiliation[c]{Department of Physics, Niigata University,\\
Niigata, Niigata 950-2181, Japan}

\affiliation[d]{Yukawa Institute for Theoretical Physics, Kyoto University,\\
Kyoto, Kyoto 606-8502, Japan}

\affiliation[e]{J-PARC Branch, KEK Theory Center, Institute of Particle and Nuclear Studies, KEK,\\
Tokai, Ibaraki 319-1106, Japan}

\emailAdd{kanaya@ccs.tsukuba.ac.jp}

\abstract{We study the finite-temperature critical point of QCD in the heavy-quark region by a scaling study of the Binder cumulant on large lattices.
Extending our previous study at $N_t=4$, we perform simulations on $N_t=6$ and 8 lattices with spatial volumes up to the aspect ratio $LT=N_s/N_t=18$ and 15 ($N_s=108$ and 120), respectively, to determine the critical point in the thermodynamic limit with a high precision. 
To enable simulations with large spatial volumes, we adopt the hopping parameter expansion combined with a method to effectively incorporate high order terms of the expansion.
The reliability of the method is confirmed by examining the effect of high order terms.
Using the results of the critical point at $N_t=4$, 6, and 8, we also attempt a preliminary continuum extrapolation of the critical point in physical units.}

\FullConference{The 41st International Symposium on Lattice Field Theory (LATTICE2024)\\
 28 July - 3 August 2024\\
Liverpool, UK\\}

\begin{document}
\maketitle

\section{Introduction}
\label{sec:intro}

In the Columbia plot in which the nature of the finite-temperature deconfinement transition in $2+1$ flavor QCD is summarized as function of the degenerate $ud$ and the $s$ quark masses, $m_{ud}$ and $m_s$, the first-order regions are expected around the light quark limit ($m_{ud} = m_s =0$) and in the heavy quark region (large $m_{ud}$ and $m_s$). 
While the physical point is located in the crossover region, thermodynamic properties around the physical point may be affected by scaling due to nearby critical points (CPs).
On the edges of the first-order regions as well as in the two-flavor chiral limit $m_{ud}=0$, we expect CPs.
Recent lattice studies suggest that the first-order region in the light-quark side is quite narrow if it exists~\cite{light1,light2,light3}. 
Because the CP in the light-quark side turned out to be more distant from the physical point than previously considered, it is important to study the influence of the CP in the heavy-quark side.
In this study, we focus on the CP in the heavy-quark region.

A powerful method in determining the location of CP is the finite-size scaling (FSS) analysis of the Binder cumulant~\cite{binder} assuming an approximate dominance of the leading singularity.
However, previous studies suggest that this requires rather large spatial lattices~\cite{cuteri,whot-itagaki,whot-kiyohara}.
We thus perform simulations with spatially large lattices.
We also adopt the reweighting method to continuously vary coupling parameters, as required by the Binder cumulant analysis.

We measure the spatial lattice size $L=N_sa$ around the critical temperature $T \sim T_c$ by the aspect ratio $LT = N_s/N_t$, where $N_s$ and $N_t$ are the spatial and temporal extent of the lattice, respectively, $a$ is the lattice spacing, and $T=1/(N_ta)$ is the temperature.
We first studied the heavy-quark QCD at $N_t=4$ on lattices with $LT=6$-12 ($N_s=24$-48)~\cite{whot-kiyohara}. 
In this report, we extend the study to $N_t=6$~\cite{whot-ashikawa} and 8~\cite{whot-sugawara} lattices with $LT$ up to $LT=18$ and 15 ($N_s=108$ and 120), respectively.

\section{Setup}
\label{sec:setup}

Our lattice action is a combination of the plaquette gauge action
\begin{align}
  S_{\rm g} = -6 N_s^3 N_t\beta \hat{P}
\end{align} 
and the standard Wilson quark action, where $N_s$ and $N_t$ are the spatial and temporal lattice sizes, and $\hat{P}$ is the plaquette.
Integrating out the quarks, the effective action is given by
\begin{align}
  S_{\rm eff} = S_{\rm g} - \sum_{f=1}^{N_{\rm f}} \ln{\rm det}M(\kappa_f),
\end{align} 
where $M$ is he Wilson quark kernel given by
\begin{align}
  M_{xy} (\kappa_f) = \delta_{xy} - \kappa_f B_{xy},
  \;\;\;\;
  B_{xy}
  =  \sum_{\mu=1}^4 \left[ (1-\gamma_{\mu})\,U_{x,\mu}\,\delta_{y,x+\hat{\mu}} + (1+\gamma_{\mu})\,U_{y,\mu}^{\dagger}\,\delta_{y,x-\hat{\mu}} \right],
  \label{eq:M}
\end{align} 
with $\kappa_f=1/(2am_f+8)$ the hopping parameter for the $f$th flavor with the bare quark mass $m_f$, and $B_{xy}$ the hopping term from $y$ to $x$.
For simplicity, we consider the case of degenerate $N_{\rm f}$ flavors in the following, although generalization to non-degenerate cases is straightforward.

\subsection{Hopping parameter expansion}
\label{sec:HPE}

In the heavy-quark region $\kappa\ll1$, we may adopt the hopping parameter expansion (HPE):
\begin{align}
    S_{\rm eff} = S_{\rm g} - N_{\rm f} N_s^3 N_t \sum_{n=1}^\infty \big( \hat W(n) + \hat L(N_t,n) \big) \kappa^n ,
    \label{eq:SeffWL}
\end{align}
where $\hat W(n)$ and $\hat L(N_t,n)$ are contributions from length-$n$ closed trajectories of $B_{xy}$ without and with windings along the temporal direction, respectively.
Here, we assume that the spatial extent is sufficiently large such that the effects of spatial windings are negligible. 
We refer to the terms included in $\hat W(n)$ and $\hat L(N_t,n)$ as the Wilson loops and the Polyakov-loop-type (PLT) loops, respectively.

To the lowest nontrivial order (LO), we have contributions proportional to the Plaquette $\hat{P}$ and the Polyakov-loop $\hat\Omega$:
\begin{align}
    S_{\rm LO} = - 96 N_s^3 N_t N_{\rm f} N_{\rm c} \kappa^4\, \hat{P} 
    - N_s^3 \lambda \, {\rm Re}\hat\Omega,
    \;\;\;\; 
    \lambda = 2^{N_t+2} N_{\rm f} N_{\rm c} \kappa^{N_t},
    \label{eq:LO}
\end{align}
where $N_{c}=3$ for QCD, and $\hat{P}$ and $\hat\Omega$ are normalized such that they become unity in the weak-coupling limit.
To the next-to-leading order (NLO), we have contributions proportional to the length-six Wilson loops with the factor $\kappa^6$ and bent PLT loops of length $N_t+2$ with the factor $\kappa^{N_t+2}$.
See refs.~\cite{whot-wakabayashi,whot-ashikawa} for details.

\subsection{Convergence of HPE}
\label{sec:convergence}

\begin{figure}[t]
 \centering
 \includegraphics[width=0.48\textwidth, clip]{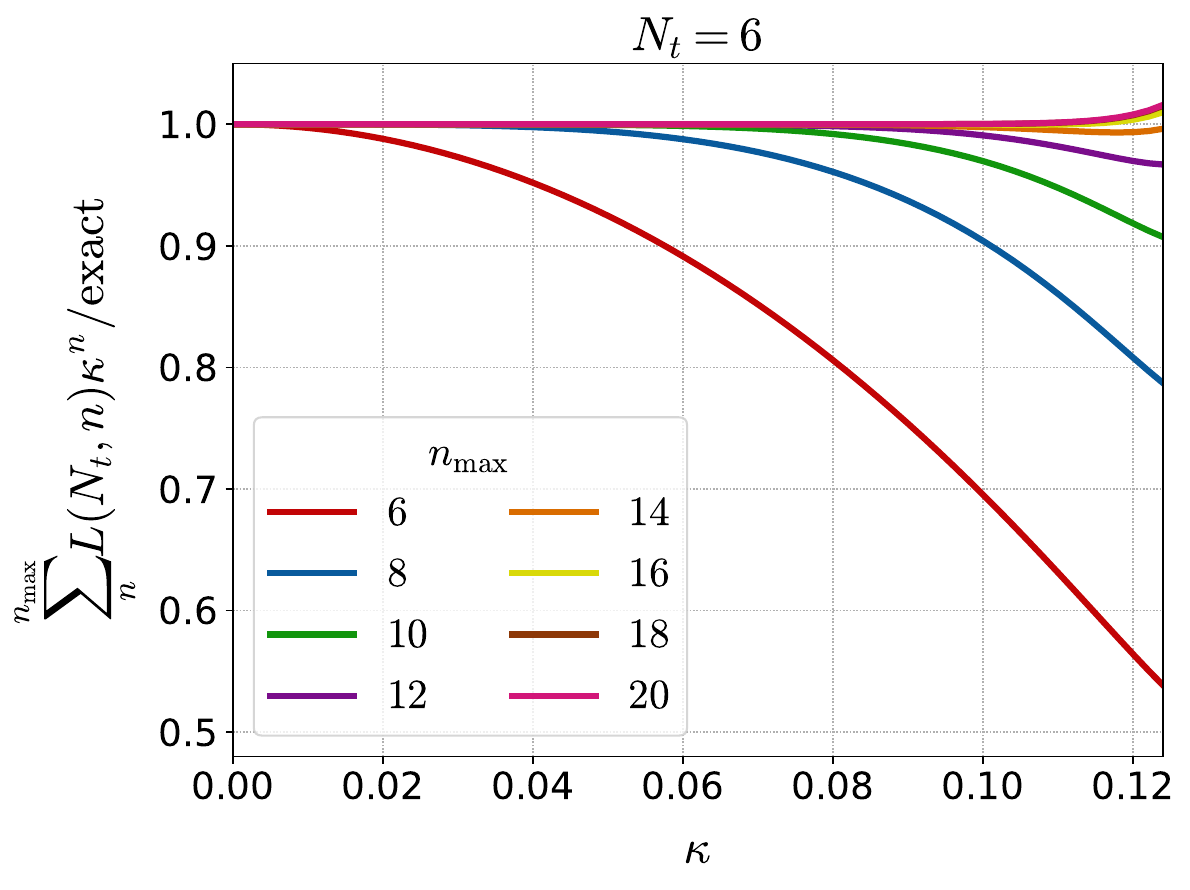}
 \hspace{2mm}
 \includegraphics[width=0.47\textwidth, clip]{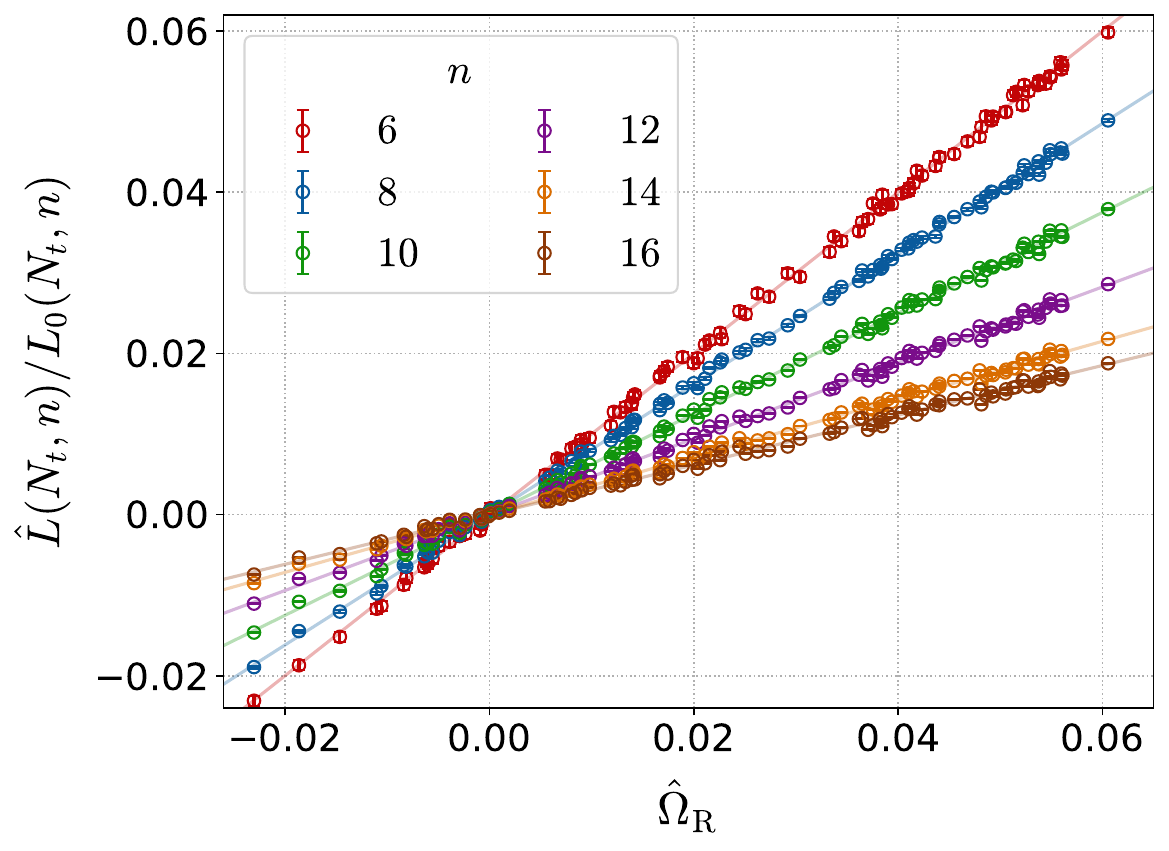}
 \caption{
   {\bf Left}: Relative deviation from the exact value due to truncation of the HPE for the PLT contribution in the effective action at $N_t=6$~\cite{whot-ashikawa}. 
   {\bf Right}: Scatter plot of $n$th order PLT terms of the HPE {it vs.} the LO Polyakov-loop, observed on an $N_t=6$ lattice near the CP~\cite{whot-ashikawa}.
 }
 \label{fig:1}
\end{figure}

Because $\kappa=1/(2m_qa+8)$, the HPE worsens as we approach the continuum limit $a\to0$.
In accordance with this, $\kappa_c$ becomes larger as $N_t$ increases, and we need more and more high-order terms of the HPE to study the CP. 

In Refs.~\cite{whot-wakabayashi,whot-ashikawa}, we studied the convergence of the HPE around the heavy-quark limit.
In the left panel of Fig.~\ref{fig:1}, we show the relative deviation from the true value due to truncation of the HPE. 
This is the result of the PLT loop terms in the effective action at $N_t=6$ in the worst convergent case~\cite{whot-ashikawa}. 
The results at $N_t=4$ and 8 as well as those for the contribution of Wilson loops are similar~\cite{whot-wakabayashi,whot-ashikawa}.
From the figure we find that, around the CP at $N_t=4$, $\kappa_c=0.0603(4)$~\cite{whot-kiyohara} for $N_{\rm f}=2$, the effective action in the LO approximation (red curve) may have an error of 10\% maximally, while the NLO (blue curve) is guaranteed to be fairly accurate.
Around $\kappa_c=0.08769(7)(^{+11}_{-0})$~\cite{whot-ashikawa} at $N_t=6$, the LO and NLO approximations may have error of 25 and 7\%, respectively, while the NNLO is quite accurate.
Around $\kappa_c=0.09024(46)$~\cite{whot-sugawara} at $N_t=8$, we need NNLO and higher orders to achieve a good accuracy.

\section{Method}
\label{sec:method}

\subsection{NLO simulation}
\label{sec:NLO}

\begin{figure}[t]
 \centering
 \includegraphics[width=0.9\textwidth, clip]{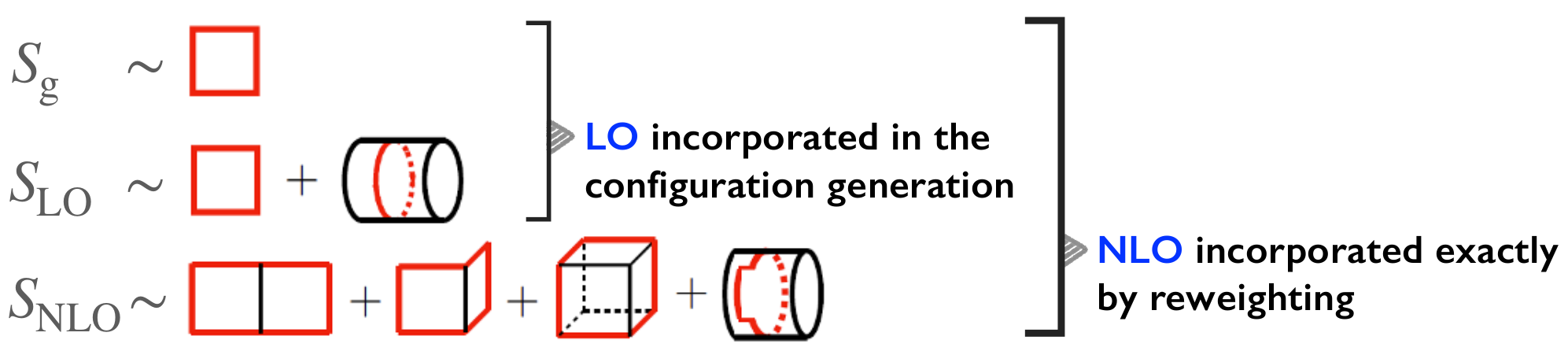}
 \caption{
   Simulation of heavy-quark QCD incorporating LO and NLO terms of the HPE~\cite{whot-kiyohara}.
 }
 \label{fig:0}
\end{figure}

As illustrated in Fig.~\ref{fig:0}, we generate gauge configurations according to the action
\begin{align}
    S_{\rm g} + S_{\rm LO} = - 6 N_s^3 N_t \beta^* \, \hat{P} 
    - N_s^3 \lambda \, {\rm Re}\hat\Omega,
    \;\;\;\; 
    \beta^* = \beta + 16N_{\rm f} N_{\rm c} \kappa^4.
    \label{eq:g+LO}
\end{align}
This action can be simulated by a pseudo-heatbath algorithm with over-relaxation, which can be efficiently implemented on vector and parallel computers~\cite{whot-kiyohara}.
We then incorporate the NLO effects exactly by the reweighting method: 
\begin{align}
  \langle \hat O \rangle_{\rm{NLO}}
= \frac{
    \langle \hat{O} e^{-S_{\rm NLO}} \rangle_{\rm LO}
  }{
    \langle e^{-S_{\rm NLO}} \rangle_{\rm LO}
  }.
\end{align}
We also adopt the reweighting method to continuously vary the coupling parameters around the CP.

The total simulation cost is comparable to that of quenched QCD simulations. 
We note that the overlap problem of the reweighting method is largely resolved by taking the LO effects in configuration generation~\cite{whot-kiyohara}.
These were essential for our study of first-order transition with large lattices.

In \cite{whot-kiyohara}, we studied the CP in heavy-quark QCD at $N_t=4$ adopting this NLO simulation.
This approximation truncating the HPE after the NLO will be sufficient to determine the CP at $N_t=4$. 
However, at $N_t \ge 6$, we need to incorporate NNLO and higher order effects. 
We do this with an effective method discussed in the next subsection.

\subsection{Effective incorporation of NNLO and higher order terms of HPE}
\label{sec:effective}

\begin{figure}[t]
 \centering
 \includegraphics[width=0.49\textwidth, clip]{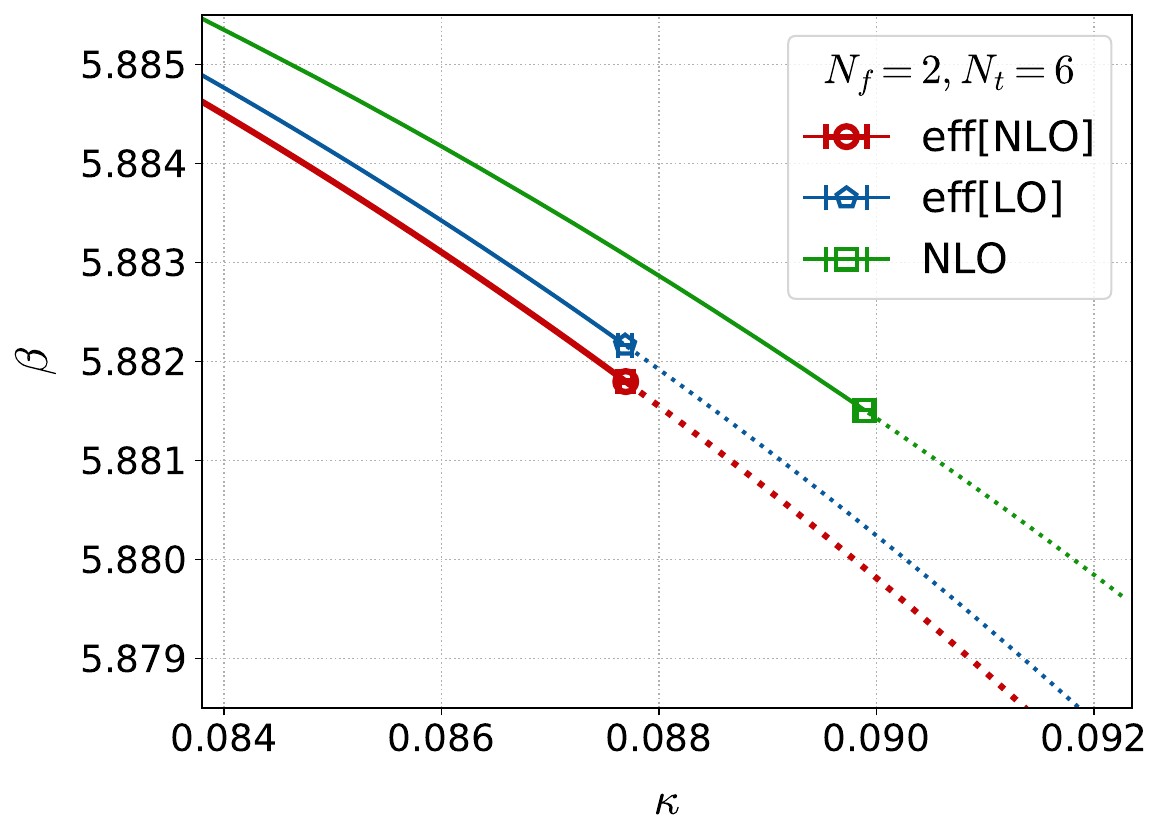}
 \hspace{3mm}
 \includegraphics[width=0.46\textwidth, clip]{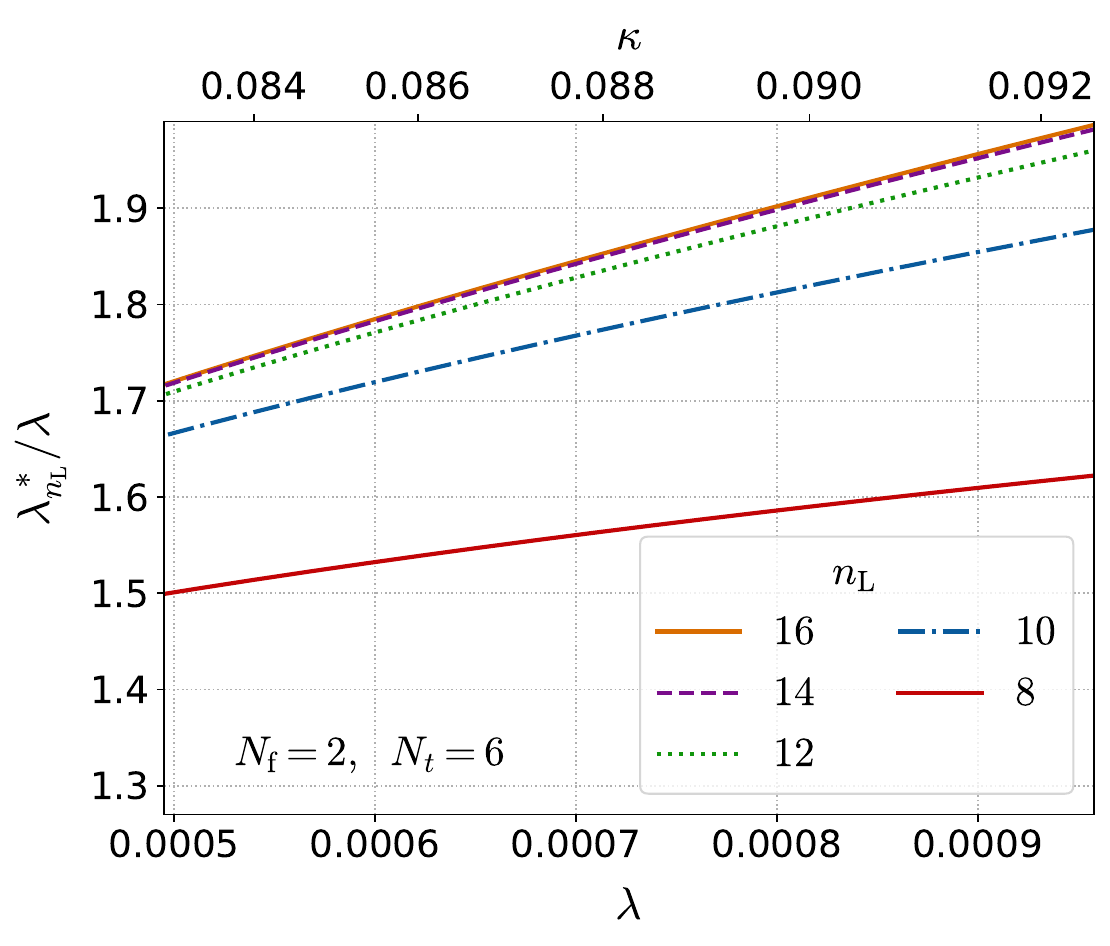}
 \caption{
   {\bf Left}: Phase diagram of two-flavor QCD in the $(\beta,\kappa)$ plane at $N_t=6$~\cite{whot-ashikawa}. The full and dotted lines represent the first-order transition line and the crossover line, while the symbols at the end of the first-order lines represent the CP. The green, blue, and red lines show the result of the NLO analysis discussed in Sec.~\ref{sec:NLO}, the eff[LO] method, and the eff[NLO] method discussed in Sec.~\ref{sec:effective}, respectively.
   {\bf Right}: Effective coupling of the Polyakov-loop term in the effective action as function of the truncation order $n_L$ in the eff[LO] method, obsereved on an $N_t=6$ lattice near the CP~\cite{whot-ashikawa}.
 }
 \label{fig:2}
\end{figure}

Our basic observation is the strong linear correlation among different order terms of HPE~\cite{whot-wakabayashi}.
In the right panel of Fig.~\ref{fig:1}, we show the scatter plot of PLT terms of the effective action, observed at $N_t=6$ near the CP.
This strong linear correlation suggests an approximation to replace high-order terms by rescaled low-order term, where the rescaling factors are given in terms of the slopes in this figure.
Wilson-loop type terms show weaker but similar correlation.

In \cite{whot-wakabayashi}, we proposed to incorporate the effects of NLO and higher-order terms by shifting the LO couplings $\beta^*$ and $\lambda$ in Eq.(\ref{eq:g+LO}) ({\bf eff[LO] method}).
Alternatively, because it is easy to incorporate the NLO effects exactly as discussed in Sec.~\ref{sec:NLO}, we may incorporate the NNLO and higher order terms by shifting the NLO couplings ({\bf eff[NLO] method})~\cite{whot-ashikawa}.
Because the latter method is exact up to the NLO, and also because the correlation is stronger with smaller order differences, the eff[NLO] method should be better than the eff[LO] method.

Let us test these methods consulting the final phase diagram obtained at $N_t=6$~\cite{whot-ashikawa} shown in the left panel of Fig.~\ref{fig:2}.
In the $(\beta,\kappa)$ plane, the first-order deconfinement transition (full lines) in the heavy-quark region terminates at the CP when we decrease the quark mass by increasing $\kappa$, and turns into crossover shown by dotted lines.
See Sec.~\ref{sec:results} for precise definition of these lines.
The green line is the result of the NLO study discussed in Sec.~\ref{sec:NLO}, and the blue and red lines are the results of the eff[LO] and eff[NLO] methods, respectively. 
We note that the blue and red lines deviate from the green line, meaning that the NNLO and higher orders are important to determine the CP at $N_t\ge6$ in this precision.
On the other hand, the red and blue lines are close to each other.
The difference between the red and blue lines is mainly whether the NLO effects are treated exactly or effectively. The closeness of them indicates that our effective method is working well in incorporating the NLO term.

We also studied the influence of vary high order terms in our effective methods~\cite{whot-ashikawa}. 
In the right panel of Fig.~\ref{fig:2}, we show the effective LO coupling $\lambda^*$ that incorporates higher-order terms up to the $n_L$th order in the eff[LO] method.
We find that the results become stable when $n_L$ is larger than 14 in this case.
We have confirmed similar stability of $\beta^*$ in the eff[LO] method as well as the effective NLO couplings in the eff[NLO] method when the truncation order is sufficiently large.

\section{Results}
\label{sec:results}

Identifying ${\rm Re}\hat\Omega$ with the magnetization in the Z(2) spin system, we study the Binder cumulant~\cite{binder}
\begin{align}
    B_4 = \frac{\langle({\rm Re}\hat\Omega)^4\rangle_c}{\langle({\rm Re}\hat\Omega)^2\rangle_c^2}+3
\end{align}
along the transition/crossover line on lattices with various spatial sizes,
where the transition/crossover line is defined as the minimum position of $B_4$ at each $\kappa$.\footnote{We also examine alternative definitions of the transition/crossover line -- as the maximum position of $\langle({\rm Re}\hat\Omega)^2\rangle_c$, and as the zero position of $\langle({\rm Re}\hat\Omega)^3\rangle_c$. 
We confirm that all three definitions lead to the same transition/crossover line in the large $LT$ limit around the CP we studied. 
See Fig.~5 of \cite{whot-ashikawa} and Fig.~8 of \cite{whot-kiyohara}. 
The crossover line may show dependence on the definitions in regions far from the CP.
We note that the conventional definition using the maximum of $\langle({\rm Re}\hat\Omega)^2\rangle_c$, {\it i.e.} the susceptibility peak, has the strongest dependence on $LT$ among these three definitions, on both the transition and the crossover sides.}
When the leading term of the FSS dominates, $B_4$ should be independent of $LT$ at the CP.

\subsection{$N_t=4$}
\label{sec:Nt4}

\begin{figure}[t]
 \centering
 \includegraphics[width=0.47\textwidth, clip]{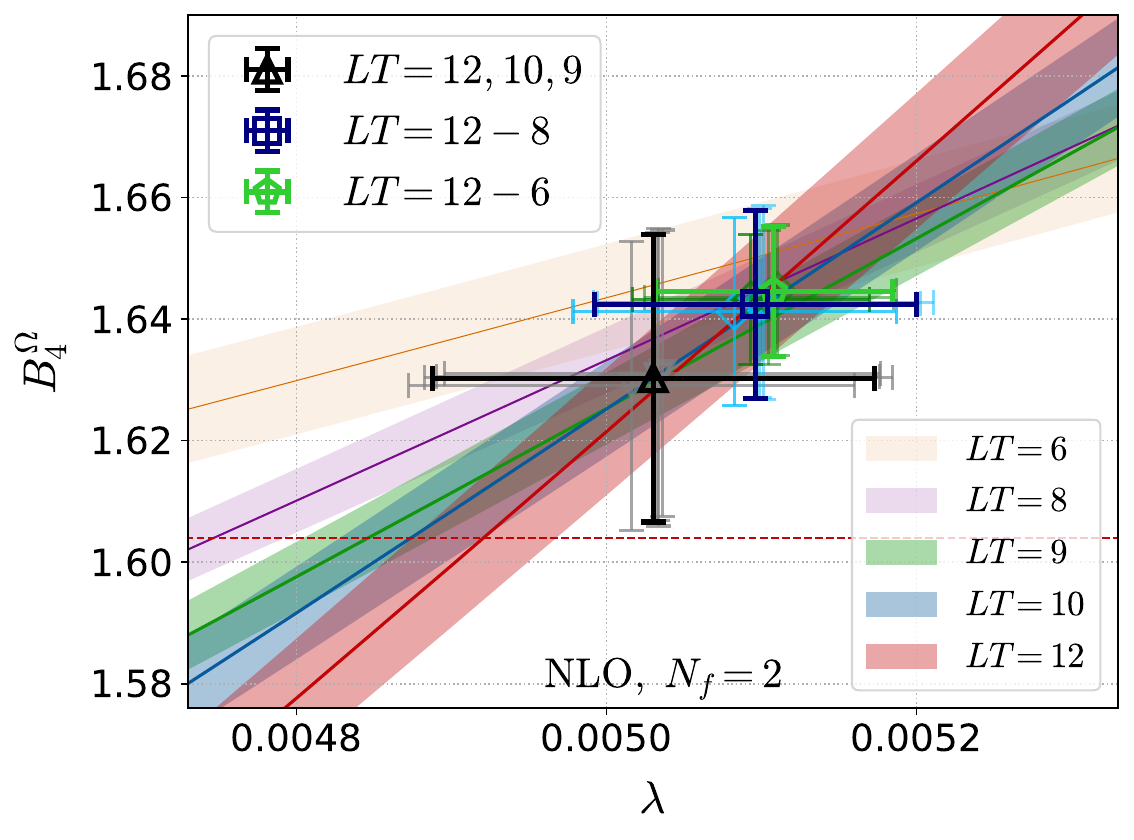}
 \hspace{3mm}
 \includegraphics[width=0.47\textwidth, clip]{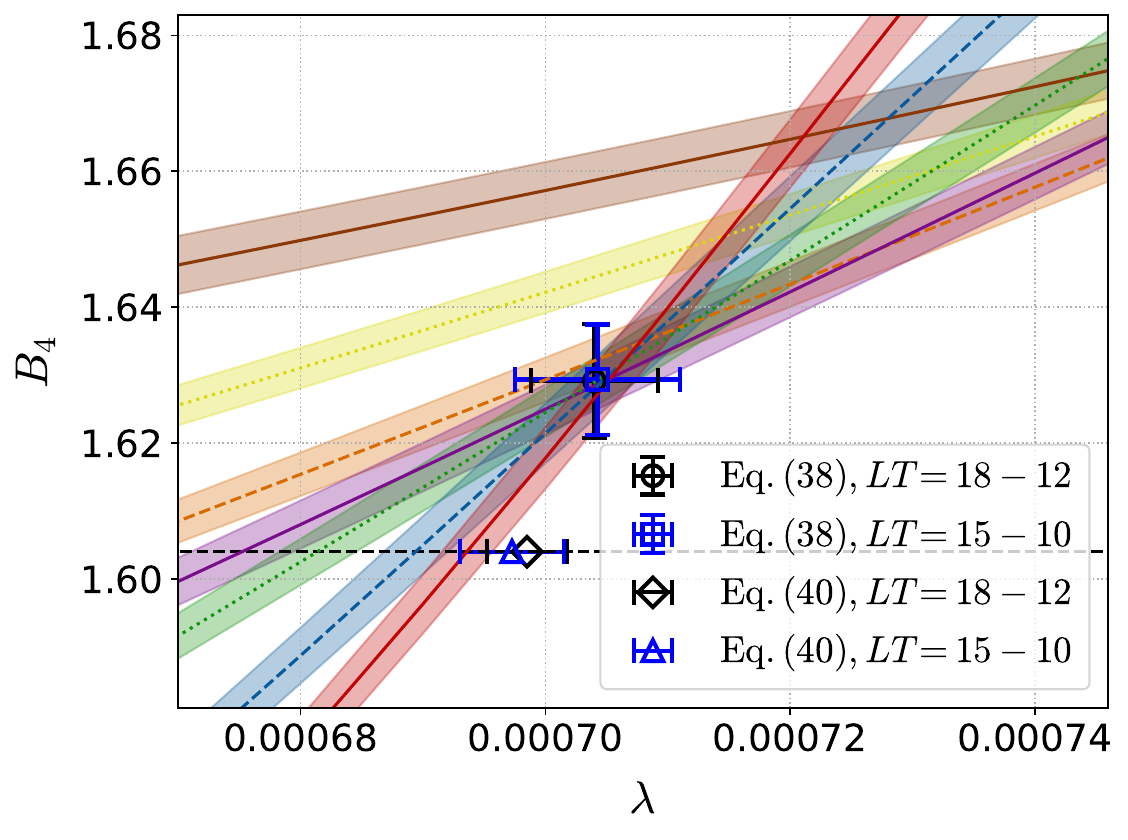}
 \caption{Binder cumulant of the Polyakov loop along the transition line, obtained with various spatial volumes.
   {\bf Left}: Results at $N_t=4$~\cite{whot-kiyohara}. 
   {\bf Right}: Results at $N_t=6$~\cite{whot-ashikawa}. The symbols with both vertical and horizontal error bars are the results of conventional FSS fits identifying the Polyakov loop as the magnetization of the Z(2) spin model, while the symbols with horizontal error bar only are the results of FSS fits in which possible contamination of the energy-like operator is taken into account.
 }
 \label{fig:3}
\end{figure}

We first study the case of $N_t=4$ on lattices with the aspect ratio $LT=N_s/N_t=6$-12 ($N_s=24$-48)~\cite{whot-kiyohara}.
Our results of $B_4$ are shown in the left panel of Fig.~\ref{fig:3}.
We find that $LT\ge9$ is required to get a stable crossing of $B_4$ to the present precision.

Performing a FSS fit using $LT=9$-12, we obtain the critical exponent $\nu=0.614(48)(3)$ and the crossing height $b_4=1.630(24)(2)$, which are consistent with the expected Z(2) values 0.630 and 1.604 within $1\sigma$.
For the CP, we obtain $\lambda_c=0.00503(14)(2)$ that corresponds to $\kappa_c=0.0603(4)$ for $N_{\rm f}=2$.
Because the flavor-dependence is analytically known in the HPE, it is straightforward to translate these results for general cases such as the $2+1$ flavor QCD.

\subsection{$N_t=6$}
\label{sec:Nt6}

We now extend the study to $N_t=6$ simulating lattices with $LT=6$-18 ($N_s=36$-108)~\cite{whot-ashikawa}. 
Our results of $B_4$ are shown in the right panel of Fig.~\ref{fig:3}.

\begin{figure}[t]
 \centering
 \includegraphics[width=0.49\textwidth, clip]{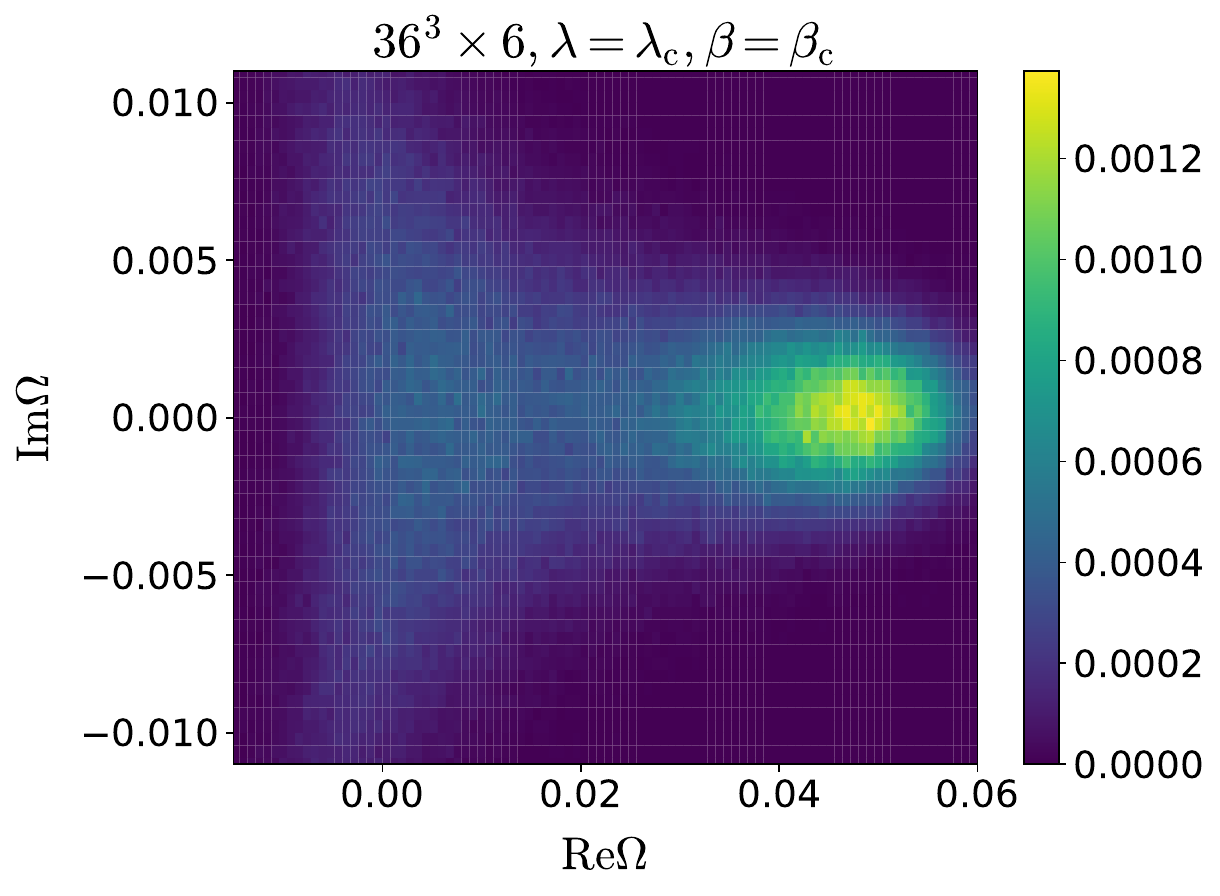}
 \hspace{2mm}
 \includegraphics[width=0.44\textwidth, clip]{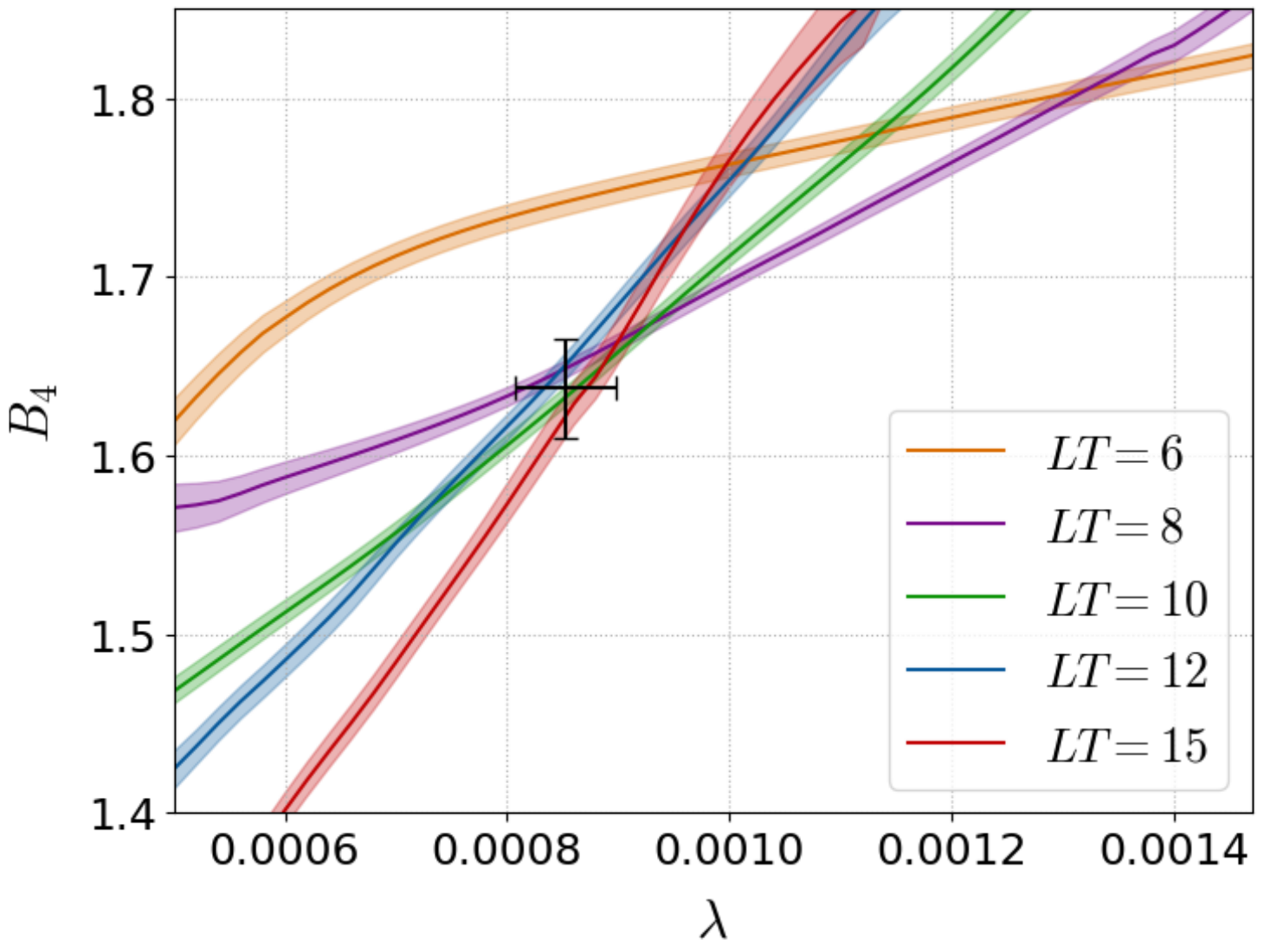}
 \caption{
 {\bf Left}: Two-dimensional histogram of $\hat{\Omega}$ on the complex plane at the critical point at $N_t=6$ and $LT=6$~\cite{whot-ashikawa}. 
 {\bf Right}: Preliminary results of the Binder cumulant obtained at $N_t=8$~\cite{whot-sugawara}. The symbol with both vertical and horizontal error bars is the crossing point estimated by a FSS fit using $LT=10$-15.
 }
 \label{fig:4}
\end{figure}

Compared to the $N_t=4$ case, we note that the violation of the FSS at small $LT$ becomes greater on the finer lattice. 
To understand the origin of the violation of the FSS at
small $LT$, we study the distribution of $\hat{\Omega}$ near the CP. Result for $LT=6$ is shown in the left panel of Fig.~\ref{fig:4}. 
We note that the distribution around the peak at ${\rm Re}\hat{\Omega}\approx0$ extends toward large $|{\rm Im}\hat\Omega|$, suggesting the remnant of the Z(3) center symmetry in the heavy-quark limit. 
The asymmetry of the distribution around the two peaks becomes weaker as $LT$ becomes larger, while it is visible even at $LT=15$.
This asymmetry of the two peaks may be causing the violation of the FSS based on the Z(2) universality around the CP.

By an FSS fit using $LT=12$-18, we obtain $\nu=0.627(19)(5)$ and $b_4=1.6297(84)(6)$, as shown by the symbols with both vertical and horizontal error bars in the figure.
Although the critical exponent is consistent with the Z(2) value, the crossing height turned out to be more than $2\sigma$ away.

This motivated us to try an alternative FSS fit in which possible contamination of the energy-like operator in ${\rm Re}\hat\Omega$ is taken into account.
Because the full six-parameter fit turned out to be unstable, we perform fits fixing the critical exponents and $b_4$ to their Z(2) values.
We find that the fits work well with acceptable $\chi^2/{\rm dof}$ when $LT\ge10$.
We thus conclude that our data obtained on spatially large lattices are consistent with the Z(2) scaling.
In the right panel of Fig.~\ref{fig:3}, the result of the CP is shown by the symbols at $b_4=1.604$ with horizontal error bar only.

Taking the result of this FSS fit using $LT=12$-18 as the central value, we obtain $\kappa_c=0.08769(7)(^{+11}_{-0})$ for $N_{\rm f}=2$, where the second bracket is the systematic error estimated from the difference with the FSS fit disregarding the contamination of the energy-like operator. 
This result is consistent with the result 0.0877(9) of the two-flavor QCD study using $LT=4$-7~\cite{cuteri}, while the error is significantly suppressed in our analysis.

\subsection{$N_t=8$}
\label{sec:Nt8}

We also extend the study to $N_t=8$ simulating lattices of $LT=6$-15 ($N_s=48$-120)~\cite{whot-sugawara}. 
Our preliminary results of $B_4$ are shown in the right panel of Fig.~\ref{fig:4}. 

Although the crossing points became more scattered, overall features are similar to the $N_t=4$ and 6 cases.
By a FSS fit using $LT=10$-15, we obtain $\nu=0.72(27)$ and $b_4=1.638(28)$, which are consistent with the Z(2) values within $1\sigma$ though the errors are large. 
By fixing $\nu$ to the Z(2) value, we obtain $b_4=1.637(24)$, confirming a stability of the crossing point.
The CP locates at $\kappa_c=0.09024(46)$ for $N_{\rm f}=2$,
to be compared with a previous result 1.1135(8) from the two-flavor QCD study using mainly $LT=4$-6 (in part 7 and 10 also)~\cite{cuteri}.

\section{Summary}
\label{sec:summary}

We determined the critical point in finite-temperature heavy-quark QCD by a finite-size scaling study of the Binder cumulant.
We found that quite large spatial lattices are required to obtain the leading finite-size scaling signal clearly.
In the heavy-quark region, such simulations with large spatial volumes are enabled by the hopping parameter expansion combined with an effective method to incorporate high orders.
Performing simulations at $N_t=4$-8, we made a precise determination of the critical point.

Using previous results of the pseudo-scalar meson mass $m_{\rm PS}$ at $T=0$~\cite{cuteri,itagaki}, we express the results of the critical point in terms of physical observables.
We obtain
\begin{align}
    &m_{\rm PS}^{\rm (CP)}/T_c = 16.30(3) &N_t=4 \;\;{\rm using}\;\; LT=9{\rm -}12,\;\,
    \\
    &m_{\rm PS}^{\rm (CP)}/T_c = 18.07(2)(^{+11}_{-0}) &N_t=6 \;\;{\rm using}\;\; LT=10{\rm -}18,
    \\
    &m_{\rm PS}^{\rm (CP)}/T_c = 17.2(2) &N_t=8 \;\;{\rm using}\;\; LT=10{\rm -}15,    
\end{align}
at the CP for $N_{\rm f}=2$.
We find that the lattice-spacing dependence is quite small up to $N_t=8$ in this combination, suggesting that these values are not far from the continuum limit.

Our method should work at least up to $N_t$ around 10.
Determination of the CP on finer lattices is important to draw a more definite conclusion about the continuum limit.
Applications to finite-density cases are also easy in the HPE.
Extensions of the study to these directions are underway.

\section*{Acknowledgment}

We thank Yasumichi Aoki, Atsushi Kiyohara, Hiroshi Suzuki, Takashi Umeda, and Naoki Wakabayashi for useful discussions. 
This work was supported in part by JSPS KAKENHI (No.~JP19H05598, No.~JP20H01903, No.~JP21K03550, No.~JP22K03593, No.~JP22K03619, No.~JP23H04507, No.~JP24K07049), and by the Center for Gravitational Physics and Quantum Information (CGPQI) at Yukawa Institute for Theoretical Physics.
Numerical simulations were made in part under the HPCI System Research projects (Project ID: hp200013, hp200089, hp210012, hp210039, hp220020, hp220024), the Research proposal-based use at the Cybermedia Center, Osaka University, and the Multidisciplinary Cooperative Research Program of the Center for Computational Sciences, University of Tsukuba.
This work is based in part on the Bridge++ code~\cite{bridge}.


\begin{thebibliography}{99}

\bibitem{light1}
Y.\ Kuramashi, Y.\ Nakamura, H.\ Ohno, and S.\ Takeda, 
\emph{Nature of the phase transition for finite temperature $N_{\rm f}=3$ QCD with nonperturbatively $O(a)$ improved Wilson fermions at $N_t=12$,}
\href{https://doi.org/10.1103/PhysRevD.101.054509}
{\emph{Phys.\ Rev.\ D} \textbf{101} (2020) 054509} 
[{\tt hep-lat/2001.04398}].

\bibitem{light2}
O.\ Philipsen, 
\emph{Lattice constraints on the QCD chiral phase transition at finite temperature and baryon density,}
\href{https://doi.org/10.3390/sym13112079}
{\emph{Symmetry} \textbf{13} (2021) 2079} 
[{\tt hep-lat/2111.03590}].

\bibitem{light3}
L.\ Dini, P.\ Hegde, F.\ Karsch, A.\ Lahiri, C.\ Schmidt, and S.\ Sharma, 
\emph{The chiral phase transition in three-flavor QCD from lattice QCD,}
\href{https://doi.org/10.1103/PhysRevD.105.034510}
{\emph{Phys.\ Rev.\ D} \textbf{105} (2022) 034510} 
[{\tt hep-lat/2111.12599}].

\bibitem{binder}
K.\ Binder, 
\emph{Finite size scaling analysis of Ising model block distribution functions,}
\href{https://doi.org/10.1007/BF01293604}
{\emph{Z.\ Phys.\ B} \textbf{43} (1981) 119}. 

\bibitem{cuteri}
F.\ Cuteri, O.\ Philipsen, A.\ Sch\"on, and A.\ Sciarra, 
\emph{Deconfinement critical point of lattice QCD with $N_{\rm f}=2$ Wilson fermions,}
\href{https://doi.org/10.1103/PhysRevD.103.014513}
{\emph{Phys.\ Rev.\ D} \textbf{103} (2021) 014513} 
[{\tt hep-lat/2009.14033}]. 

\bibitem{whot-itagaki}
S.\ Ejiri, S.\ Itagaki, R.\ Iwami, K.\ Kanaya, M.\ Kitazawa, A.\ Kiyohara, M.\ Shirogane, and T.\ Umeda, 
\emph{End point of the first-order phase transition of QCD in the heavy quark region by reweighting from quenched QCD,}
\href{https://doi.org/10.1103/PhysRevD.101.054505}
{\emph{Phys.\ Rev.\ D} \textbf{101} (2020) 054505} 
[{\tt hep-lat/1912.10500}].

\bibitem{whot-kiyohara}
A.\ Kiyohara, M.\ Kitazawa, S.\ Ejiri, and K.\ Kanaya, 
\emph{Finite-size scaling around the critical point in the heavy quark region of QCD,}
\href{https://doi.org/10.1103/PhysRevD.104.114509}
{\emph{Phys.\ Rev.\ D} \textbf{104} (2021) 114509} 
[{\tt hep-lat/2108.00118}].

\bibitem{whot-wakabayashi}
N.\ Wakabayashi, S.\ Ejiri, K.\ Kanaya, and M.\ Kitazawa, 
\emph{Scope and convergence of the hopping parameter expansionin finite-temperature quantum chromodynamics with heavy quarks around the critical point,}
\href{https://doi.org/10.1093/ptep/ptac019}
{\emph{Prog.\ Theor.\ Exp.\ Phys.} \textbf{2022} (2022) 033B05} 
[{\tt hep-lat/2112.06340}].

\bibitem{whot-ashikawa}
R.\ Ashikawa, M.\ Kitazawa, S\. Ejiri, and K.\ Kanaya, 
\emph{High-precision analysis of the critical point in heavy-quark QCD at $N_t = 6$,}
\href{https://doi.org/10.1103/PhysRevD.110.074508}
{\emph{Phys.\ Rev.\ D} \textbf{110} (2024) 074508} 
[{\tt hep-lat/2407.09156}]. 

\bibitem{whot-sugawara}
H.\ Sugawara et al., for the WHOT-QCD Collaboration, in preparation.

\bibitem{itagaki}
S.\ Itagaki et al., for the WHOT-QCD Collaboration, unpublished.

\bibitem{bridge}
S.\ Ueda et al., 
\emph{Development of an object oriented lattice QCD code 'Bridge++',}
\href{https://doi.org/10.1088/1742-6596/523/1/012046}
{\emph{J.\ Phys.\ Conf.\ Ser.} \textbf{523} (2014) 012046};
\href{http://bridge.kek.jp/Lattice-code/}
{\tt http://bridge.kek.jp/Lattice-code/}.

\end{thebibliography}
\end{document}